\documentclass[english]{article}
\usepackage{caption}
\usepackage{appendix}
\usepackage{authblk}
\usepackage{float}
\usepackage{slashbox}
\usepackage[utf8]{inputenc}

\title{$QCD_3$ Dualities and the F-Theorem}
\author{Adar Sharon}
\affil{Department of Particle Physics and Astrophysics, Weizmann Institute of Science, Rehovot, Israel}
\date{}

\makeatletter
\let\runtitle\
\let\runauthor\
\makeatother

\usepackage{hyperref}
\hypersetup{
	colorlinks,
	citecolor=blue,
	filecolor=blue,
	linkcolor=blue,
	urlcolor=blue,
	linktocpage=true
}

\usepackage{amsmath}
\usepackage{mathptmx}
\usepackage{amsthm}
\usepackage{amssymb}
\usepackage{amsfonts}
\usepackage{cancel}
\usepackage{verbatim}
\DeclareMathAlphabet{\mathcal}{OMS}{cmsy}{m}{n}

\usepackage[margin=1in]{geometry}

\usepackage[compress,numbers]{natbib}

\usepackage{graphicx}

\usepackage{fancyhdr}
\usepackage[rgb,dvipsnames]{xcolor}
\newcommand\todo[1]{\textcolor{red}{(#1)}}

\newcommand{\munu}{{\mu\nu}}
\newcommand{\munurho}{{\mu\nu\rho}}

\newcommand{\dnu}{{\partial_\nu}}

\newcommand{\psib}{\bar{\psi}}

\newcommand\mc[1]{\mathcal{#1}}
\newcommand{\tr}{\text{Tr}}

\newcommand{\grd}{\delta}
\newcommand{\grD}{\Delta}
\newcommand{\gre}{\epsilon}

\newcommand{\grf}{\phi}

\newcommand{\grl}{\lambda}
\newcommand{\grs}{\sigma}

\newcommand{\grr}{\rho}

\newcommand{\gry}{\psi}

\begin{document}
\maketitle

\begin{abstract}There has recently been a surge of new ideas and results for 2+1 dimensional gauge theories. We consider a recently proposed duality for 2+1 dimensional QCD, which predicts a symmetry-breaking phase. Using the F-theorem, we find bounds on the range of parameters for which the symmetry-breaking phase (and the corresponding duality) can occur. We find exact bounds for an $ SU(2) $ gauge theory, and approximate bounds for an $SU(N) $ gauge theory with $ N>2 $.\end{abstract}

\tableofcontents

\appendixtitleon
\appendixtitletocon
\newpage

\section{Introduction}
Dualities in 2+1 dimensional theories have been gaining increasing attention recently, partially due to progress in localization of 2+1$ d $ supersymmetric partition functions \cite{Hsin:2016blu,Seiberg:2016gmd,Benini:2017dus,Aharony:2015mjs,Karch:2016sxi,Kachru:2016rui,Kachru:2016aon,Jafferis:2011ns,Kapustin:2011vz}. Specifically, the low energy phase diagrams of various generalizations of $ QCD_3 $ have recently been discussed \cite{Komargodski2017,Gomis:2017ixy}, leading to some non-trivial results at strong coupling.
 
In this paper, we study the dualities and phase diagrams discussed in \cite{Komargodski2017}. In particular, we discuss the symmetry-breaking phase conjectured to appear for strongly-coupled $ QCD_3 $, where the Chern-Simons level $ k $ is small and the theory has $ N_f $ fermions such that $ 2k<N_f<N_\star(N,k) $ for some unknown $ N_\star $. The purpose of this paper will be to find some bound on the value of $ N_\star(N,k) $.

A useful method to test when the symmetry-breaking phase can appear was discussed in \cite{Grover2012}. There, the method used the F-theorem \cite{Klebanov2011, Casini:2012ei,Jafferis:2011zi,Myers:2010tj,Casini:2011kv} (see \cite{Pufu2016} for a review) in order to constrain the RG flow from $ QED_3 $ to a chiral symmetry-breaking phase. In this paper we use a similar method in order to constrain the RG flows discussed in \cite{Komargodski2017}.

The general idea is the following. Define the F-coefficient of a 2+1$ d $ theory as $ F=-\ln|Z_{S^3}| $, where $Z_{S^3}$ is the partition function of the theory on the 3-sphere. The F-theorem is the conjecture that this quantity is monotonically decreasing along RG flows\footnote{This definition of the F-theorem is subtle since the sphere partition function is well defined only at RG fixed points. A more precise statement of the F-theorem is that if we can flow from CFT$_1$ to CFT$ _2 $ then their F-coefficients obey $ F_1\geq F_2 $.} (this is the 2+1$ d $ analog of the c-theorem and the a-theorem in 1+1$ d $ and 3+1$ d $ respectively \cite{Zamolodchikov:1986gt,Cardy:1988cwa,Komargodski:2011vj}). The intuition usually associated with these theorems is that $ F $ (along with $ c,a $ in their corresponding dimensions) measures the number of degrees of freedom in the theory, which should intuitively decrease as we decrease the energy scale of our theory. Similar theorems exist for higher dimensions and even for non-integer dimensions \cite{Dowker:2017cqe,Giombi:2014xxa}.

Now, suppose we would like to test whether some theory $\mc{A}$ with F-coefficient $ F_{\mc{A}} $ can flow to some IR theory with F-coefficient $ F_{IR} $. A simple test would be to find the F-coefficients for the two theories, and check whether they obey the conjectured inequality $ F_{\mc{A}}\geq F_{IR} $. Unfortunately, F-coefficients are not always simple to calculate. Furthermore, a complication arises when calculating F-coefficients in gauge theories. If theory $\mc{A}$ is that of $ N_f $ fermions coupled to a gauge group (say $ SU(N) $ or $ U(1) $), naively one would have wanted to set the F-coefficient at the UV fixed point to be the sum of the F-coefficients of $ N_f $ free fermions and of the free gauge fields. Unfortunately, the free gauge field contribution diverges in the UV fixed point. This can be seen through an explicit calculation for $ U(1) $ gauge theories \cite{Klebanov2011_2}, and is related to the fact that free Maxwell theory is not conformal in  2+1$d$. Since the F-coefficient of the UV theory diverges, we find that we cannot use the F-theorem to constrain its RG flow.

Instead, in this paper we will be using the following trick \cite{Grover2012}. We will take a supersymmetric (SUSY) theory and calculate the F-coefficient at its IR fixed point. We will then show that we can flow from this fixed point to the theory $\mc{A}$, which proceeds to flow to our IR theory. This solves our problem, since we no longer need to calculate UV F-coefficients for gauge theories. Additionally, since our theory is supersymmetric, we can also try to use localization in order to calculate the F-coefficient. Using this new RG flow, the F-theorem states $ F_{SUSY}\geq F_{IR} $, and so if this inequality is not obeyed, then we can conclude that the theory A \textbf{cannot} flow to the IR theory, leading to a constraint on the RG flow. Of course, our bound will be better the "closer" the SUSY theory is to theory $\mc{A}$ in RG space.

\begin{figure}
	\centering
	\includegraphics[width=0.6\linewidth]{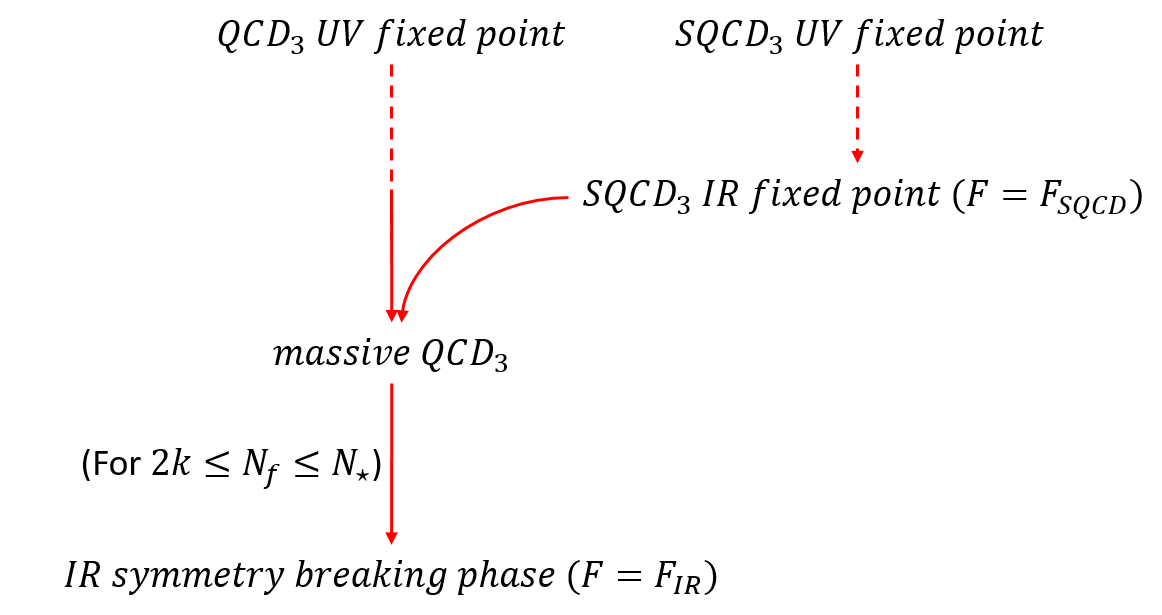}
	\captionsetup{width=.8\linewidth}
	\caption{The RG flow we will be testing using the F-theorem. Dashed lines represent points which are at an infinite distance in RG space (corresponding to a diverging F-coefficient), and arrows represent RG flows. The F-theorem now states that if theory $\mc{A}$ can flow to theory $\mc{B}$ then we must have $ F_\mc{A}\geq F_\mc{B} $. In particular, we must have $F_{SQCD}\geq F_{IR}$. Thus, if for $N_f=N_f^{bound}$ we have $F_{SQCD}= F_{IR}$, then we must have $ N_\star\leq N_f^{bound} $.}
	\label{fig:rg-flow}
\end{figure}

In our paper, the SUSY theory will be 2+1$d $ $ \mc{N}=2 $ $ SU(N)_{\tilde{k}} $ with $ N_f $ fundamental chiral multiplets ($SQCD_3$), the theory $\mc{A}$ will be $ SU(N)_{k} $ with $ N_f $ fundamental fermions ($QCD_3$), and the IR theory will be the symmetry-breaking phase (SB) with a non-linear sigma model (NLSM). Since the F-coefficient for the IR theory is easily calculated, the bulk of this paper will be devoted to the calculation of the SUSY F-coefficient using localization and F-maximization \cite{Jafferis2010}. The RG flow described above is summarized in Figure \ref{fig:rg-flow}.

Using this method, we successfully find bounds $ N_f^{bound} $ such that $ N_\star\leq N_f^{bound} $. For $ SU(2) $ we calculate the bounds both numerically and using a saddle-point approximation for large $ N_f $. We find that the two methods agree almost exactly for all values of $ k $. Specifically, we find $ N_f^{bound}=13 $ for $ k=0 $ (this result can be compared to recent results using lattice simulations \cite{Karthik:2018nzf}, which for $ SU(2) $ with $ k=0 $ found $ N_\star\leq8 $). For a general $ SU(N) $ gauge theory with $ N>2 $ we calculate approximate bounds using a saddle point approximation. Since the saddle-point approximation agrees with the numerical results for $ SU(2) $, it is safe to assume that this approximation is very good for $ N>2 $ as well. Interestingly, we find that for all gauge theories $ SU(N) $ with $ N\geq2 $, we can never completely rule out a symmetry-breaking window in the theory (even at very large $ k $). This might be due to the fact that symmetry breaking can happen even at very large $ k $, although the more probable explanation is that the bounds obtained using this method are just not stringent enough to exclude this window at large $ k $.

The outline of this paper is as follows. In Section \ref{background} we review the proposal in \cite{Komargodski2017}, and study the RG flow from $ SQCD_3 $ to $ QCD_3 $ and then to the IR symmetry-breaking phase. We also discuss the calculation of the F-coefficients in these theories using localization. In Section \ref{Nstar For SU(2)} we calculate a bound on $ N_\star $ for an $ SU(2) $ gauge theory. Finally, in Section \ref{Nstar for SU(N)} we generalize these results for a general $ SU(N) $ gauge theory.

\section{Background}\label{background}

\subsection{Phases of $ QCD_3 $}

We quickly review the proposal in \cite{Komargodski2017}. Consider an $ SU(N)_k $ gauge theory coupled to $ N_f $ fermions in the fundamental representation in 2+1$ d $ (which we call $ QCD_3 $). Here, we adopt the notation in \cite{Komargodski2017} for the Chern-Simons level, resulting in time reversal acting as $ k\rightarrow-k $ in $ QCD_3 $ (note that in this notation, $ k$ is half integer when $ N_f $ is odd and integer when $ N_f $ is even). Throughout this paper, $ k $ will always denote the $ QCD_3 $ Chern-Simons level as defined above, while $ \tilde{k} $ will denote the corresponding Chern-Simons level for SUSY theories.

One could ask what are the low-energy phases of this theory as a function of the fermion masses $ m $. It turns out that different phases appear in different regimes of the theory:
\begin{itemize}
	\item For $ 2k\geq N_f $, the two phases of the theory are $ SU(N)_{k\pm N_f/2}$, with a single phase transition between them \cite{Hsin:2016blu,Seiberg:2016gmd,Benini:2017dus,Aharony:2015mjs,Karch:2016sxi,Kachru:2016rui,Kachru:2016aon,Murugan:2016zal,Radicevic:2016wqn,Karch:2016aux,PhysRevB.95.205137,Peskin:1977kp,Dasgupta:1981zz,Son:2015xqa,Intriligator:1996ex,deBoer:1996mp,Aharony:1997bx,Aharony:1997gp,Giveon:2008zn,Kapustin:2011gh,Willett:2011gp,Benini:2011mf,Aharony:2011ci,Aharony:2013dha,PhysRevB.89.235116,PhysRevX.6.031026,PhysRevB.94.245107,Park:2013wta,Aharony:2013kma,Sezgin:2002rt,Klebanov:2002ja,Giombi:2011ya,Aharony:2011jz,Giombi:2011kc,Maldacena:2011jn,Aharony:2012nh,Giombi:2012ms,Aharony:2012ns,Jain:2013py,Jain:2013gza,Jain:2014nza,Inbasekar:2015tsa,Minwalla:2015sca,Gur-Ari:2016xff}.
	\item For  $  2k < N_f < N_\star (N,k) $, where the theory is strongly coupled, the following phase diagram was proposed in \cite{Komargodski2017}: When the fermion masses are large, the phases are still $ SU(N)_{k\pm N_f/2} $. However, for small masses a new phase appears, which is a NLSM with a Wess-Zumino term. For $ N>2 $ its target space is $ \frac{U(N_f)}{U(N_f/2+k)\times U(N_f/2-k)} $, while for $ N=2 $ its target space is $ \frac{Sp(N_f)}{Sp(N_f/2+k)\times Sp(N_f/2-k)} $. This is consistent with some conjectured dualities at the transition points. There are two such critical points, One where the dual theory is $U\left(\frac{N_f}{2}+k\right)_{-N}$ with $N_f$ bosons,, and another where the dual theory is $U\left(\frac{N_f}{2}-k\right)_{N}$ with $N_f$ bosons.
	\item for $ N_f\geq N_\star $ the exact behavior is not known, apart from the fact that at large enough $ N_f $ the phases should once again contain only Chern-Simons terms, as in the regime $ 2k\geq N_f $ \cite{Appelquist1989}.
\end{itemize}

Here, $ N_\star $ is some upper bound on the symmetry-breaking (SB) phase. This bound must exist, since in the limit $ N_f\rightarrow\infty $, the theory does not develop dynamical masses for the fermions and thus symmetry breaking cannot occur \cite{Appelquist1989}. Intuitively, one might say that for very large $ N_f $, the theory becomes weakly coupled (we will see explicitly that this is true by calculating the dimension of the chiral multiplets and the F-coefficients for these theories for large $ N_f $).

The purpose of this paper is to find a bound on $ N_\star $. That is, we attempt to constrain the values of the parameters $ N_f,N,k $ for which symmetry breaking can occur. In order to find this bound, we shall use the method described in \cite{Grover2012}, where a similar bound for $ QED_3 $ was calculated. Let us describe this method in the present context. 

Our method will rely on the F-theorem. Consider the theory of a 2+1$ d $ $ \mathcal{N}=2 $ supersymmetric $ SU(N)_{\tilde{k}} $ gauge theory with $ N_f $ chiral multiplets (which we call $ SQCD_3 $). By adding appropriate deformations, we can flow from $ SQCD_3 $ to a non-supersymmetric $ SU(N)_k $ gauge theory with $ N_f $ fermions ($ QCD_3 $), and from there the proposed flow in \cite{Komargodski2017} is to the IR symmetry-breaking phase. We then calculate the F-coefficient of the SUSY theory $ F_{SUSY} = F_{SQCD} $, and the F-coefficient of the symmetry-breaking IR theory $ F_{IR}=F_{SB} $. The F-theorem then states that the proposal is valid only when $ F_{SQCD} \geq F_{IR} $ (with $ F_{SQCD},F_{IR} $ functions of $ N_f,N,k $). Now, if we define $ N_f^{bound}(N,k) $ as the value of $ N_f $ for which we have $ F_{SQCD} = F_{IR} $, then for any $N_f$ such that $ N_f> N_f^{bound} $, we must have $ F_{SQCD} < F_{IR} $. This is in conflict with the F-theorem, and so we conclude that we must have $ N_\star \leq N_f^{bound}  $  (this is summarized in Figure \ref{fig:rg-flow}).

This paper will focus on finding values of the bound $ N_f^{\text{bound}}(N,k) $, that is, we find the value of $ N_f $ for which $ F_{SUSY}=F_{IR} $. In order to find these values, one must compute $ F_{SUSY}=F_{SQCD} $ and $ F_{IR}=F_{SB} $. The calculation of the F-coefficient for the IR symmetry-breaking phase is simple, while the calculation of $ F_{SQCD} $ will be much more complicated. Both will be discussed in the next sections.	

\subsection{RG Flow}\label{RG flows}
We now discuss the RG flow from $ SQCD_3 $ with CS level $ \tilde{k} $ to $ QCD_3 $ with CS level $ k $, and from there to the IR symmetry-breaking phase. Since a time reversal transformation takes $ \tilde{k}\rightarrow-\tilde{k} $, it suffices to consider $ \tilde{k}\geq0 $. We shall further restrict ourselves to the case $ \tilde{k}\geq N $, in order to avoid problems with SUSY breaking and runaways\footnote{This assumption is very natural in our context, since we will find that $ \tilde{k}\geq N $ corresponds to the QCD CS level $ k $ obeying $ k\geq0 $.} \cite{Intriligator2013a}.  We begin by writing down the content of our theory explicitly. We write down the Lagrangian for canonical R-charge $ \grD=\frac{1}{2} $ (this is both for simplicity and because we will show that corrections from F-maximization can be neglected in this work). $ SQCD_3 $ consists of an $ \mc{N}=2 $ vector multiplet and $ N_f $ chiral multiplets, and its Lagrangian on $ S^3 $ consists of three parts \cite{Jafferis2010,Schwarz:2004yj,Hosomichi2010}:
\begin{align*}
\mc{L}_{CS} &= \frac{\tilde{k}}{4\pi} \tr\left[\gre^\munurho \left(A_\mu\dnu A_\grr - \frac{2i}{3}A_\mu A_\nu A_\grr\right)-\bar{\lambda}\grl + 2D \grs\right]\\
\mc{L}_{YM} & =\frac{1}{2g^2}\tr \left[\frac{1}{2}F_\munu F^\munu+D_\mu \sigma D^\mu \sigma+(D+\grs)^2+i\bar{\lambda}\cancel{D}\lambda+i\bar{\lambda}[\sigma,\lambda]-\frac{1}{2}\bar{\lambda}\grl\right]\\
\mc{L}_M & =D_\mu \phi^\dagger D^\mu \phi +\frac{3}{4}\phi^\dagger\phi+i\overline{\psi}\cancel{D}\psi + F^\dagger F+\phi^\dagger \sigma^2 \phi+i\phi^\dagger D\phi-i\psib \sigma\psi+i\phi^\dagger \bar{\lambda} \psi-i\psib \lambda\phi
\end{align*}
here we have suppressed the flavor and color indices, and we have set the radius of $ S^3 $ to $ 1 $. In terms of SUSY multiplets, the above consists of:
\begin{itemize}
	\item An $\mc{N}=2$ vector multiplet: a gauge field $ A_\mu $ (with $ F_\munu $ its field strength), a real boson $ \grs $, a Dirac fermion $ \grl $ and an auxiliary real boson $ D $.
	\item $ N_f $ copies of an $\mc{N}=2$ fundamental chiral multiplet: a complex boson $ \grf_i $, a Dirac Fermion $ \gry_i $ and an auxiliary complex boson $ F_i $ for $ i=1,...,N_f $.
\end{itemize}

Let us discuss the RG flow from $ SQCD_3 $ to $ QCD_3 $, following \cite{Jain2013}. We start by integrating out the auxiliary fields $ F,D $. Next, we add a large negative mass for the gaugino $ \grl $, and when integrating it out we obtain a shift in the CS level\footnote{The shift is by $ N $ since the gaugino is a complex fermion in the adjoint of $ SU(N) $. The sign of the shift is due to the fact that the gaugino has negative mass. Note that the CS term in the Lagrangian gives the gaugino a negative mass when $ \tilde{k}>0 $. It is important here that we have	 $ \tilde{k}\geq N $, allowing us to integrate it out.} $ \tilde{k}\rightarrow \tilde{k}-N $. Finally, we add masses to the bosons $ \grs,\grf $ such that the $ U(N_f) $ symmetry is unbroken, and integrate them out as well.  This process will result in an $ SU(N)_{\tilde{k}-N} $ gauge theory with $ N_f $ fermions. The fermions $ \gry_i $ will be massive, but since the deformations we introduced respect the $ U(N_f) $ symmetry, all of the fermions $ \gry_i $ will have the same mass $ m $. The results obtained in \cite{Jain2013} also show that by modifying the deformations we introduce, one can change the fermion mass $ m $ in the IR at will (the results in \cite{Jain2013} were obtained using large $ N $ calculations, but it is safe to assume that even for finite $ N $ we can reach small enough values of the masses $ m $ so that we are in the symmetry-breaking phase\footnote{In order to flow to the correct range of masses $m$, we must assume that there is no phase transition as a function of the boson mass $m_s$ for $m_s\neq0$. This provides us with the dimensionless parameter $m_s/g^2$ which we can tune to make sure that $m/g^2$ is small enough after integrating out the scalars.}).

To summarize, by tuning the deformations we introduce we can flow from $ SQCD_3 $ to $ QCD_3 $ such that the masses of the fermions $ m $ are small enough so that we are in the symmetry-breaking phase. If we begin with some $ \tilde{k} $ in $ SQCD_3 $, we end up with QCD with $ k=\tilde{k}-N $. Importantly, we thus have 
\begin{equation}
\tilde{k} = k + N
\label{eq:ktilde vs k}
\end{equation}
which relates the SUSY CS level $ \tilde{k} $ with the non-SUSY CS level $ k $. This equation will be very useful for us, since $ \tilde{k} $ will appear in our localization calculations, but we will mostly be interested in the corresponding value of $ k $.

\subsection{IR F-coefficient}	
	
We calculate the F-coefficient for the symmetry-breaking phase. Start with $ N>2 $ and assume that the symmetry breaking described in \cite{Komargodski2017} takes place. The IR is a NLSM with target space $\frac{U\left(N_{f}\right)}{U\left(\frac{N_{f}}{2}+k\right)\times U\left(\frac{N_{f}}{2}-k\right)}$ and with a WZ term, and so in the deep IR the theory contains only free massless bosons\footnote{This point is subtle, since the fact that the NLSM is compact can cause the F-coefficient to diverge. For example, if the target space was $S^1$, this would have prevented the appearance of the conformal coupling (that is proportional to $\phi^2 R$, with $\phi$ the boson and $R$ the curvature) in the IR, causing the F-coefficient to diverge. To remedy this, we can add an irrelevant operator in the UV which flows to the conformal coupling in the IR. Since the operator is irrelevant this will not change the RG flow, but will make the F-coefficient finite. Thus we can take the IR F-coefficient to be the sum over F-coefficients of independent conformal bosons.}. The number of bosons is the number of broken generators: $$N_{bosons}=N_f^2-\left[\left(\frac{N_f}{2}+k\right)^2+\left(\frac{N_f}{2}-k\right)^2\right]=2\left(\frac{N_f^2}{4}-k^2\right)$$
The F-coefficient of a 2+1$ d $ free boson was computed in \cite{Klebanov2011}, and the result is $ F_\grf \approx0.0638 $ (see also Appendix \ref{Some F-coefficients}). Thus, the F-coefficient of the IR theory is\footnote{The fact that we can use the F-coefficient of a free massless boson here is subtle, since the bosons we have here are compact.}

\begin{equation}
F_{IR}^{N>2}=N_{bosons}\cdot F_\grf=2\left(\frac{N_f^2}{4}-k^2\right)F_\grf
\approx 0.1276\left(\frac{N_f^2}{4}-k^2\right)
\label{eq:FIR}
\end{equation}

A similar calculation for the $ N=2 $ case with target space $\frac{Sp\left(N_{f}\right)}{Sp\left(\frac{N_{f}}{2}+k\right)\times Sp\left(\frac{N_{f}}{2}-k\right)}$ leads to
\begin{equation}
F_{IR}^{N=2}=4\left(\frac{N_f^2}{4}-k^2\right)F_\grf
\approx 0.2552\left(\frac{N_f^2}{4}-k^2\right)
\label{eq:FIR N=2}
\end{equation}
The next step is calculating the F-coefficient for $ SQCD_3 $. To do this, we use localization.

\subsection{UV F-coefficient: Localization of 2+1d $SU(N)$ Gauge Theories}

We now turn to the calculation of the F-coefficient for the SUSY theory. The F-coefficients of supersymmetric $ U(N)_{\tilde{k}} $ gauge theories have been calculated using localization \cite{Jafferis2010,Hosomichi2010,Kapustin2009}, and they can be easily modified to describe $ SU(N)_{\tilde{k}} $ gauge theories. The sphere partition function for 2+1$ d $ $ \mc{N}=2 $ SUSY $ U(N)_{\tilde{k}} $ gauge theories with $N_{f}$ chiral multiplets in the fundamental representation and $\overline{N}_f$ chiral multiplets in the
anti-fundamental representation was computed to be

\begin{equation}
Z_{S^3}=\frac{1}{N!} \int \prod_{i=1}^N d\lambda_i e^{-i\pi \tilde{k}\lambda_i^2} \prod_{j<k} \left(2\sinh\left(\pi\left(\lambda_j-\lambda_k\right)\right)\right)^2 
\prod_{m=1}^N e^{N_fl(1-\grD+i\grl_m)}
\prod_{n=1}^{N}e^{\overline{N}_fl(1-\grD-i\grl_n)}
\label{eq:partition_func_arbitrary_delta}
\end{equation}
where $ \grD $ is the R-charge of the chiral multiplets. The function  $ l(z) $ is given by 
$$
l(z)=-z\ln(1-e^2\pi i z)+\frac{i}{2}\left(\pi z^2 + \frac{1}{\pi}Li_2(e^2\pi i z)\right)-\frac{i\pi}{12}
$$
or, equivalently, as the solution to the equation $ \partial_zl(z)=-\pi z \cot(\pi z) $ with initial condition $ l(0)=0 $ \cite{Jafferis2010}. The integral appearing in \eqref{eq:partition_func_arbitrary_delta} is an integral over the Cartan of $ U(N) $, which can be taken to be the set of diagonal matrices of the form $ \text{diag}(\grl_1,...,\grl_N) $.

How will the expression \eqref{eq:partition_func_arbitrary_delta} be modified for the group $ SU(N) $? Since the generators of $ SU(N) $ are traceless, we must also take the Cartan to be traceless, which means that the Cartan can be taken to be the set of traceless diagonal matrices. This can be achieved in \eqref{eq:partition_func_arbitrary_delta} by introducing a delta function of the form $ \grd\left(\sum_{i=1}^{N}\grl_i\right) $ into the integral. In other words, the partition function for an $ \mc{N}=2 $ $ SU(N)_{\tilde{k}} $ gauge theory is 
\begin{equation}
Z_{S^3}=\frac{1}{N!}\int \prod_{i=1}^N d\lambda_i e^{-i\pi \tilde{k} \lambda_i^2}  \prod_{j<k}(2\sinh(\pi(\lambda_j-\lambda_k)))^2 \prod_{m=1}^N e^{N_fl(1-\grD+i\grl_m)}
\prod_{n=1}^{N}e^{\overline{N}_fl(1-\grD-i\grl_n)}
\delta\left(\sum_{l=1}^N\lambda_l \right)
\label{eq:SU(N) partition function}
\end{equation}

Here we have allowed for arbitrary R-charge $ \grD $ for the chiral multiplets (for 3$ d $ $ \mc{N}=2 $ chiral multiplets, the absolute value of the R-charge $ |r| $ is equal to the dimension $ \grD $). Generically, we cannot assume that the R-charge takes the free field value $ \grD=1/2 $. In particular, one has to be careful when there exists an additional abelian flavor symmetry. If this is the case, there is no unique choice for the $ U(1)_R $ symmetry that is used to couple the theory to the curvature of $ S^3 $ \cite{Pufu2016,Jafferis2010}. In \cite{Jafferis2010} it was proposed that the correct R-charge $ \grD $ is the one for which the partition function $ |Z_{S^3}|^2 $ is minimized, and so to find the correct R-charge one must minimize $ |Z_{S^3}| $ (or equivalently maximize $ F=-\ln|Z_{S^3}| $). This method is called F-maximization. 
We will perform F-maximization in our calculations, but we will see that the corrections due to F-maximization can be neglected to the order in $ \frac{1}{N_f} $ we will be working in when using a saddle point approximation.

Finally, it is clear that the integral \eqref{eq:SU(N) partition function} is very difficult to calculate in general. For an $ SU(2) $ gauge theory we can calculate the integral and the effects due to F-maximization numerically. However, for $ SU(N) $ with $ N\geq 3 $ we will instead be using a saddle-point approximation for large $ N_f $\footnote{Note that the integral can also be calculated using a saddle point approximation for large $ \tilde{k} $. However, since we are interested in the regime where $ N_f\geq 2k $, large $ \tilde{k} $ will also result in large $ N_f $.} (this approximation was used for a $ U(N) $ gauge theory in \cite{Klebanov2011_2}). A review of the general scheme to be used appears in Appendix \ref{general idea appendix}. The bounds we find with this method will thus be approximate, and will only be valid when the resulting bound $ N_f^{bound} $ will be large (however, we will see that at least for the $SU(2)$ case they agree almost perfectly with the numerical results).

\section{$N_\star$ For $SU(2)$}\label{Nstar For SU(2)}

We start with an $ SU(2) $ gauge theory. This theory has two important simplifications compared to a general $ SU(N) $ theory - first, the fundamental and anti-fundamental representations of $ SU(2) $ are equivalent, allowing us to set $ \bar{N}_f=0 $. Second, the integral over the Cartan in the localization procedure becomes a one-dimensional integral, making it easier to calculate.

Applying these simplifications to \eqref{eq:partition_func_arbitrary_delta} we find that the integral we need to calculate is
\begin{align}
Z & =\frac{1}{N!} \int \prod_{i=1}^{N} d\lambda_i e^{-i\pi \tilde{k}\lambda_i^2} \prod_{j<k}\left(2\sinh\left(\pi\left(\lambda_{j}-\lambda_{k}\right)\right)\right)^2 \prod_{m=1}^N e^{N_fl(1-\grD+i\grl_m)}
\delta\left(\sum_{l=1}^N\lambda_l \right)=\\
&  = 2 \int dx e^{-2 i \pi \tilde{k} x^2} e^{ N_f (l(1-\grD +ix)+l(1-\grD -ix))} \sinh^2(2 \pi x)
\label{SU(2) partition function}
\end{align}
We will first calculate the integral numerically to obtain exact results for $ N_f^{bound} $. We will then redo the calculation, this time using a saddle point approximation for the integral, and compare the results for the two methods. We will find that the two methods agree almost exactly.

\subsection{Numerical Results}\label{SU(2) numerical}

We can calculate the UV F-coefficient $ F_{SUSY} $ exactly, by using F-maximization on the partition function \eqref{SU(2) partition function}. Having found $ F_{UV} $, we compare it to $ F_{IR} $ in order to find $ N_f^{bound} $. We round up the result for $ N_f^{bound} $ to an integer for convenience.

For small $ k $, the results are:
\begin{center}
	\begin{tabular}{c |c c c c c c}
		$k$ & 0 &1&2&3&4&5\\
		\hline
		$ N_f^{bound} $&13&14&15&16&17&19
	\end{tabular} 
\end{center}
For more general $ k $, we can summarize the results in the following figure:
\begin{figure}[H]
	\centering
	\includegraphics[width=0.8\linewidth]{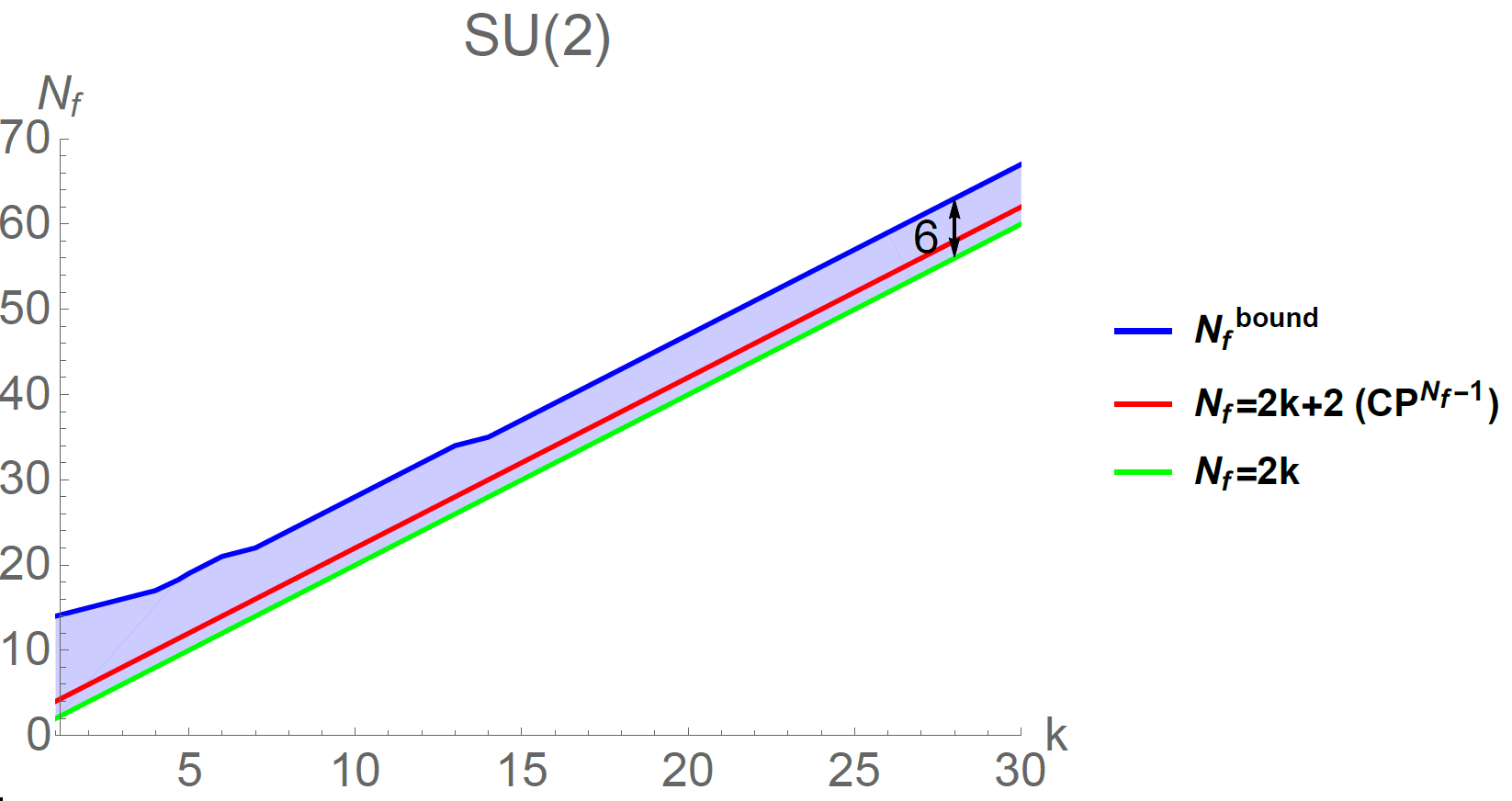}
	\captionsetup{width=.8\linewidth}
	\caption{Numerical results for $ N_f^{bound} $ for an $ SU(2) $ gauge theory. The symmetry breaking phase can appear only inside the shaded area.}
	\label{fig:SU(2)}
\end{figure}

We have plotted the lines $ N_f=2k $ (above which the symmetry-breaking phase appears), $ N_f=2k+2 $ (the first example where symmetry breaking can appear, which results in a $ \mathbb{CP}^{N_f-1} $ model in the IR) and $ N_f=N_f^{bound} $ (which was calculated numerically). The shaded area in the figure is the space of parameters in which symmetry breaking can occur. Since $ N_f^{bound} $ is only a bound, the actual space of parameters in which symmetry breaking occurs will be some subset of the area shown in the figure.

An interesting thing has happened. We expect that in the limit $ k\rightarrow
\infty $, the symmetry-breaking phase should disappear, since the theory is weakly coupled. That is, for large enough $ k $, there should not be a symmetry-breaking phase for any $ N_f $. However, the bound we have found does not force the symmetry-breaking phase to completely disappear for large k. We have found that in the limit of large $k$ and $N_f $, the symmetry-breaking phase can occur only for $ N_f $'s that obey $ 2k<N_f<2k+6 $, meaning that this method fails to eliminate a symmetry-breaking window of "size" $ 6 $.

We stress that this result does not imply that the bounds we have obtained are incorrect. This result only shows that the bounds might not be very stringent, and can definitely be improved. In particular, the closer $ F_{SQCD} $ is to $ F_{QCD} $, the better our bound will become. Unfortunately, in flowing from $ SQCD $ to $ QCD $ we have integrated out a large amount of fields (specifically, we have integrated out about $ N\cdot N_f $ fields for large $ N_f $), meaning that our $ F_{SQCD} $ might be much larger than $ F_{QCD} $. This fact may have made our bounds quite weak.

\subsection{Saddle-Point Approximation for Large $ N_f $}

We now attempt to obtain the same results using a saddle-point approximation to calculate the integral \eqref{SU(2) partition function} for large $ N_f $. To allow for $ \tilde{k} $ to also be large, we will assume a general relation of the form $ \tilde{k}=a N_f + b $ for some constants $ a,b $.

We will find that the results obtained using the saddle point approximation are in almost perfect agreement with the numerical results. This will be very good news, since it will be very difficult to obtain numerical results for a general $ SU(N) $ gauge theory. Instead, we will be using only the saddle point approximation for a general $ SU(N) $ gauge theory. Since the results agree for $ N=2 $, we assume that the saddle point approximation will generate a good approximation for $ N>2 $ as well.  

\subsubsection{F-maximization}\label{SU(2) F maximization}
We start by calculating the correction to the dimension $ \grD $ due to F-maximization, and then show that these corrections can be neglected in the F-coefficient to order $ O(1) $. 

Note that for infinite $ N_f $, when the theory is weakly coupled, we expect $ \grD $ to obtain its free field value $ \grD=1/2 $. Following \cite{Klebanov2011_2,Jafferis2010}, we can now calculate the leading-order correction to $ \grD $ by defining $ \grD=\frac{1}{2}+\frac{\grd}{N_f}+O(\frac{1}{N_f^2}) $ and calculating $ \grd $. Calculating the partition function and minimizing $ |Z_{S^3}| $, we find that the leading order correction is given by 
\begin{equation}
\grD=\frac{1}{2}-\frac{6}{(16a^2+\pi^2)N_f}+O\left(\frac{1}{N_f^2}\right)\label{eq:r charge large Nf}
\end{equation} 
where we have set $ \tilde{k}=aN_f+b $.

One can check this result by comparing it with a similar result from \cite{Jafferis2010}. There, the correction was calculated for large $ \tilde{k} $ to be $ -\frac{3(N_f-1)}{8\tilde{k}^2} $. In order to compare the results, one can take $ a\rightarrow\infty $, which corresponds to $ \tilde{k}\gg N_f\gg1 $. Equation \eqref{eq:r charge large Nf} then becomes $ \grd = -\frac{3N_f}{8\tilde{k}^2} $, in agreement with the result from \cite{Jafferis2010} to leading order in $ \tilde{k},N_f$.

We now show that F-maximization gives corrections to the F-coefficient of order $ O(1/N_f) $, and so will be ignored in the following. For $ \grD=\frac{1}{2}+\frac{\grd}{N_f}+O(\frac{1}{N_f^2}) $, we expand the exponent in \eqref{SU(2) partition function} around $ \grD=1/2 $, obtaining 
$$e^{N_f(l(1/2+ix)+l(1/2-ix))}e^{\grd(l'(1/2+ix)+l'(1/2-ix)) +O(1/N_f)}$$
Note that the integral we must calculate now in \eqref{SU(2) partition function} is the integral for $ \grD=1/2 $, i.e. without F-maximization, multiplied by $e^{\grd(l'(1/2+ix)+l'(1/2-ix)) +O(1/N_f)}$. We can now proceed with the saddle point approximation as done in Appendix \ref{general idea appendix}, and find the contribution due to this additional factor when expanding around the saddle point. Noticing that the saddle point is still at $ x=0 $ (since the function in the exponent is symmetric in $ x $) and using the fact that $ l'(1/2)=0 $, we find that its contribution will be to multiply the result by $ (1+O(1/N_f)) $. In other words, we have found that for $ \grD=\frac{1}{2}+\frac{\grd}{N_f}+O(\frac{1}{N_f^2}) $, the sphere partition function is $ Z_{\grD} = Z_{\grD=1/2}\cdot (1+O(\frac{1}{N_f})) $. We thus have $ F_{\grD}=-\ln|Z_{\grD}| = F_{\grD=1/2}+O(1/N_f) $, meaning that to order $ O(1) $ we can ignore the corrections due to F-maximization and just set $ \grD=1/2 $.

\subsubsection{Calculating $ N_\star $}

We calculate the F-coefficient \eqref{SU(2) partition function} in the saddle point approximation for large $ N_f $. We start with small $ k $. Since we found that F-maximization only affects the F-coefficient to order $ O(1/N_f) $, we can neglect it and set $ \grD=1/2 $. We can thus use the calculation of the F-coefficient for $ \grD=1/2 $ as given in Appendix \ref{SU(2)_without F_max_appendix}.

For small $ k $ and using equation \eqref{eq:ktilde vs k}, we find that we must plug in $ b=k+N,\;a=0 $ into the result from Appendix \ref{SU(2)_without F_max_appendix}, which results in
$$ F_{SQCD} = N_f \ln2 + \frac{3}{2}\ln N_f + \frac{1}{2}\ln\left(\frac{\pi}{128}\right) + O\left(\frac{1}{N_f}\right) $$
We can now find $ N_f^{bound} $ by equating this result with the IR F-coefficient given in \eqref{eq:FIR N=2}. We find an almost perfect agreement with the numerical results. In fact, in the range $ 0\leq k\leq 30 $, the numerical results and the approximation disagree only twice (and in both cases the disagreement is by 1, the lowest possible value). This agreement is partly due to the fact that $ N_f $ is an integer, and so our results for $ N_f^{bound} $ are rounded up.

We can perform a similar calculation for large $ k $. Recall that the symmetry-breaking phase occurs only for $ k<\frac{N_f}{2} $, and so by "large $ k $" we mean $ k\lesssim\frac{N_f}{2} $. We thus use Appendix \ref{SU(2)_without F_max_appendix} once more to calculate the F-coefficient, this time plugging in $ b=N-c-1 $ and $ a=\frac{1}{2} $. We obtain

$$ F_{SQCD} = N_f \ln2 + \frac{3}{2}\ln N_f + \frac{1}{2}\ln\left(\frac{(4 + \pi^2)^{3/2}}{128\pi^2}\right) + O\left(\frac{1}{N_f}\right) $$
And we can find $ N_\star $ by comparing to the IR F-coefficient \eqref{eq:FIR N=2}. In particular, we can find an analytic expression for $ N_f^{bound} $ for large enough $ N_f $ by ignoring the $ \ln N_f $ and the constant terms. The result is
$$ N_f^{bound} = 2k+5.43  + O(1/k^2)$$
We find here exactly the result we found in our numerical investigation - for large $ k$ and $N_f $, we cannot exclude a finite-sized symmetry-breaking window. We have found that in this limit, symmetry breaking can occur for $ 2k<N_f<2k+5.43 $. We note that the results here also match the numerical results almost perfectly, with some isolated cases for which they differ by 1.

\subsection{Conclusions}\label{conclusions}

Our results were summarized in Figure \ref{fig:SU(2)}. As we discussed in Section \ref{SU(2) numerical}, we have obtained a strange result for large $ k $, where we cannot exclude a finite-sized symmetry breaking window. This result was confirmed analytically.

We can now compare our results to others found in the literature. First, we can compare our results to some constraints on $ N_\star $ obtained in \cite{Komargodski2017}. In particular, it was found that $ N_\star $ must obey $ N_\star(N,k)-1\leq N_\star(N,k\pm\frac{1}{2}) $, which led to two interesting conclusions. The first is that the size of the symmetry-breaking window is maximized at $ k=0 $ (which can be clearly seen in out plot). The second is that $ N_\star $ cannot increase or decrease too fast - the average derivative as a function of $ k $ must be no more than two. Once again, we can see this in our result as well, with the slope rising asymptotically to $ 2 $ as $ k\rightarrow\infty $.

Next, we can also compare to results from lattice simulations \cite{Karthik:2018nzf}. Lattice simulations for the $SU(2)_0$ gauge theory provide strong evidence for $Sp(N_f)$ symmetry breaking for $N_f\leq 2$, and for its absence for $N_f\geq 8$. We thus conclude that $N_\star\leq 8$ when $k=0$. This should be compared to our result for $k=0$, which was $N_\star=N_f^{bound}=13$. We find that while our bounds are comparable to results from lattice simulations, they can definitely be improved.

Finally, we once again emphasize that the results found using the saddle point approximation were in excellent agreement with the numerical results. We will thus focus on the saddle point approximation when we discuss a general $ SU(N) $ gauge theory, since a numerical calculation becomes increasingly complicated when $ N>2 $.

\section{$ N_\star $ for General $SU(N)$}\label{Nstar for SU(N)}

We now attempt to find $ N_\star $ for a general $ SU(N) $ gauge theory. We will not be using numerics, and instead we will only be using a saddle point approximation. However, since we saw that this was an excellent approximation for the $ SU(2) $ gauge theory, we expect good results for $ N>2 $ as well.

\subsection{Saddle Point Approximation}

Consider the theory of $ SQCD_3 $ with only fundamental matter (that is, we set $ \overline{N}_f=0 $ once again). The partition function is given by \eqref{eq:SU(N) partition function}:
\begin{align*}
Z&=\frac{1}{N!} \int \prod_{i=1}^N d\lambda_i e^{-i\pi \tilde{k}\lambda_i^2 } 
\prod_{j<k}\left(2\sinh\left(\pi\left(\lambda_j-\lambda_k\right)\right)\right)^2
\prod_{m=1}^N e^{N_fl(1-\grD+i\grl_m)}
\grd\left(\sum_{l=1}^N\grl_l\right)=\\
&= \frac{2^\frac{2N(N-1)}{2}}{N!}\int \prod_{i=1}^N d\lambda_i e^{-i\pi \tilde{k}\lambda_i^2}
\prod_{j<k}\sinh^2\left(\pi\left(\lambda_j-\lambda_k\right)\right)
\exp\left(N_f\sum_{m=1}^N l(1-\grD+i\grl_m)\right)
\grd\left(\sum_{l=1}^N\grl_l\right)
\end{align*}
Let us find the leading order contribution in $ N_f $, assuming again $ \tilde{k}=aN_f+b $. Again, F-maximization will give us a correction to $ \grD $ of order $ \frac{1}{N_f} $, so that we expect $ \grD=\frac{1}{2}+\frac{\grd}{N_f} +... $ for some constant $ \grd $. Thus a similar proof to the one given in Section \ref{SU(2) F maximization} for an $ SU(2) $ gauge theory will show that we can neglect the corrections due to F-maximization here as well to order $ O(\frac{1}{N_f}) $\footnote{The proof here is slightly more complicated, since the saddle point is not at $ \grl_i=0 $ but at $ \grl_i=O\left(\frac{1}{N_f}\right) $. However, an argument that is similar to the proof in Section \ref{SU(2) F maximization} shows that expanding around $ \grl_i=0 $ instead of the real saddle point will still give corrections to the F-coefficient of order $ O\left(\frac{1}{N_f}\right) $.}. 

We can thus set $ \grD=\frac{1}{2} $, which gives

$$ Z_{S^3}  =\frac{2^{N(N-1)}}{N!} \int \prod_{i=1}^N d\lambda_i e^{-i\pi \tilde{k}\lambda_i^2} 
\exp\left(N_f\sum_{m=1}^N l(1/2+i\grl_m)+O\left(\frac{1}{N_f}\right)\right)
\prod_{j<k}\sinh^2\left(\pi\left(\lambda_j-\lambda_k\right)\right)
\grd\left(\sum_{l=1}^N\grl_l\right)\left(1+O\left(\frac{1}{N_f}\right)\right )
$$
We now proceed similarly to Appendix \ref{general idea appendix}. The saddle point is at $ \grl=0+O(\frac{1}{N_f}) $, and expanding the exponent around the saddle point we find
$$ Z_{S^3}  =\frac{2^{N(N-1)}}{N!} \int \prod_{i=1}^N d\lambda_i e^{-N\frac{\ln2}{2}N_f} e^{\left(-\frac{\pi ^2}{4}-i \pi  a\right) N_f \grl_i^2} 
\prod_{j<k}\sinh^2\left(\pi\left(\lambda_j-\lambda_k\right)\right)
\grd\left(\sum_{l=1}^N\grl_l\right)\left(1+O\left(\frac{1}{N_f}\right)\right )
$$
Redefining $ \grl_i\rightarrow\grl_i\sqrt{N_f} $ we find
\begin{align*}
	Z_{S^3}&  =\frac{2^{N(N-1)}e^{-\frac{\ln2}{2}NN_f}}{N! N_f^{(N-1)/2}}  \int \prod_{i=1}^N d\lambda_i  e^{\left(-\frac{\pi ^2}{4}-i \pi  a\right) \grl_i^2} 
	\prod_{j<k}\sinh^2\left(\pi\frac{\lambda_j-\lambda_k}{\sqrt{N_f}}\right)
	\grd\left(\sum_{l=1}^N\grl_l\right) \left(1+O\left(\frac{1}{N_f}\right)\right )\\
	&=\frac{2^{N(N-1)}e^{-\frac{\ln2}{2}NN_f}}{N! N_f^{(N+1)(N-1)/2}}  \int \prod_{i=1}^N d\lambda_i  e^{\left(-\frac{\pi ^2}{4}-i \pi  a\right) \grl_i^2} 
	\prod_{j<k}^N\left(\pi(\lambda_j-\lambda_k)\right)^2
	\grd\left(\sum_{l=1}^N\grl_l\right) \left(1+O\left(\frac{1}{N_f}\right)\right )
\end{align*} 
We notice that the remaining integral is independent of $ N_f $. We have thus found that at leading order in $ N_f $ we have
\begin{equation}
F_{SQCD} = -\ln|Z_{S^3}| = \frac{NN_f}{2}\ln2 + \frac{N^2-1}{2}\ln N_f + C + O\left(\frac{1}{N_f}\right)
\label{eq:SU(N)_F_coeff}
\end{equation} 
Where $ C $ is a constant given by
$$ C= -\ln \left[\frac{2^{N(N-1)}}{N!} \left| \int \prod_{i=1}^N d\lambda_i  e^{\left(-\frac{\pi ^2}{4}-i \pi  a\right) \grl_i^2} 
\prod_{j<k}^N\left(\pi(\lambda_j-\lambda_k)\right)^2
\grd\left(\sum_{l=1}^N\grl_l\right)\right|\right] $$
In particular we note that the expression \eqref{eq:SU(N)_F_coeff} reduces to the result we obtained in previous sections when one plugs in $ N=2 $.

Let us discuss the form of the F-coefficient we have obtained in equation \eqref{eq:SU(N)_F_coeff}. Consider the first term. Since the F-coefficient of a free $ \mc{N}=2 $ chiral multiplet is $\frac{\ln2}{2}$ (see Appendix \ref{Some F-coefficients}), we find that this term is just the F-coefficient of $ N\cdot N_f $ free chiral multiplets. Indeed, we could have expected this term, since our theory has $ N\cdot N_f $ chiral multiplets and it becomes weakly coupled in the large $ N_f $ limit.
Next, we see that the second term is proportional to $ \dim G $, where $ G=SU(N) $ is our gauge group. We thus recognize this term as the leading order contribution due to the gluons. Indeed, note that for pure $ \mc{N}=2 $ $ SU(N)_k $ gauge theory, the leading order term in the large-k expansion of the F-coefficient is $ \frac{N^2-1}{2}\ln |k| $. So we can think of this term as the result of integrating out the chiral multiplets when the theory is weakly coupled, leading to a shift $ k\rightarrow k\pm N_f/2 $, which for large $ N_f $ would indeed result in the second term in equation \eqref{eq:SU(N)_F_coeff} up to corrections of order $ O(1) $.

We can now find $ N_f^{bound} $ by comparing the $ F_{SQCD} $ in \eqref{eq:SU(N)_F_coeff} and $ F_{IR} $ from \eqref{eq:FIR}. We proceed just as we did for the $ SU(2) $ gauge theory in Section \ref{Nstar For SU(2)}. Let us start with small $ k $, that is, we set $ a=0 $ and $ b=k+N $. For $ N=3,4,5 $ and all $ 0\leq k\leq2 $, the results are:
\begin{center}
	\begin{tabular}{c |c c c c}
		$N$ & 3 &4&5&6\\
		\hline
		$ N_f^{bound} $&44&60&76&93
	\end{tabular} 
\end{center}

Next, we consider $ k\lesssim\frac{N_f}{2} $, that is, we set $ a=1/2 $. We can obtain analytic results by ignoring the $ \ln N_f $ and the constant terms in $ F_{SQCD} $. We find that for large $ k,N_f $ we have
$$ N_f^{bound} =  2k+5.432N  + O(1/k^2)$$
Once again, we find that our method fails to make the symmetry-breaking window completely disappear for large $ k $. Instead, the symmetry-breaking window goes to some finite size as $ k\rightarrow\infty $ (Note that the size of the window not agree with the $ N=2 $ result when plugging in $ N=2 $. This results from the fact that the IR theory is different in $ N=2 $).

\subsection{Conclusions}

We can conclude this section with the following figure:
\begin{figure}[H]
	\centering
	\includegraphics[width=0.8\linewidth]{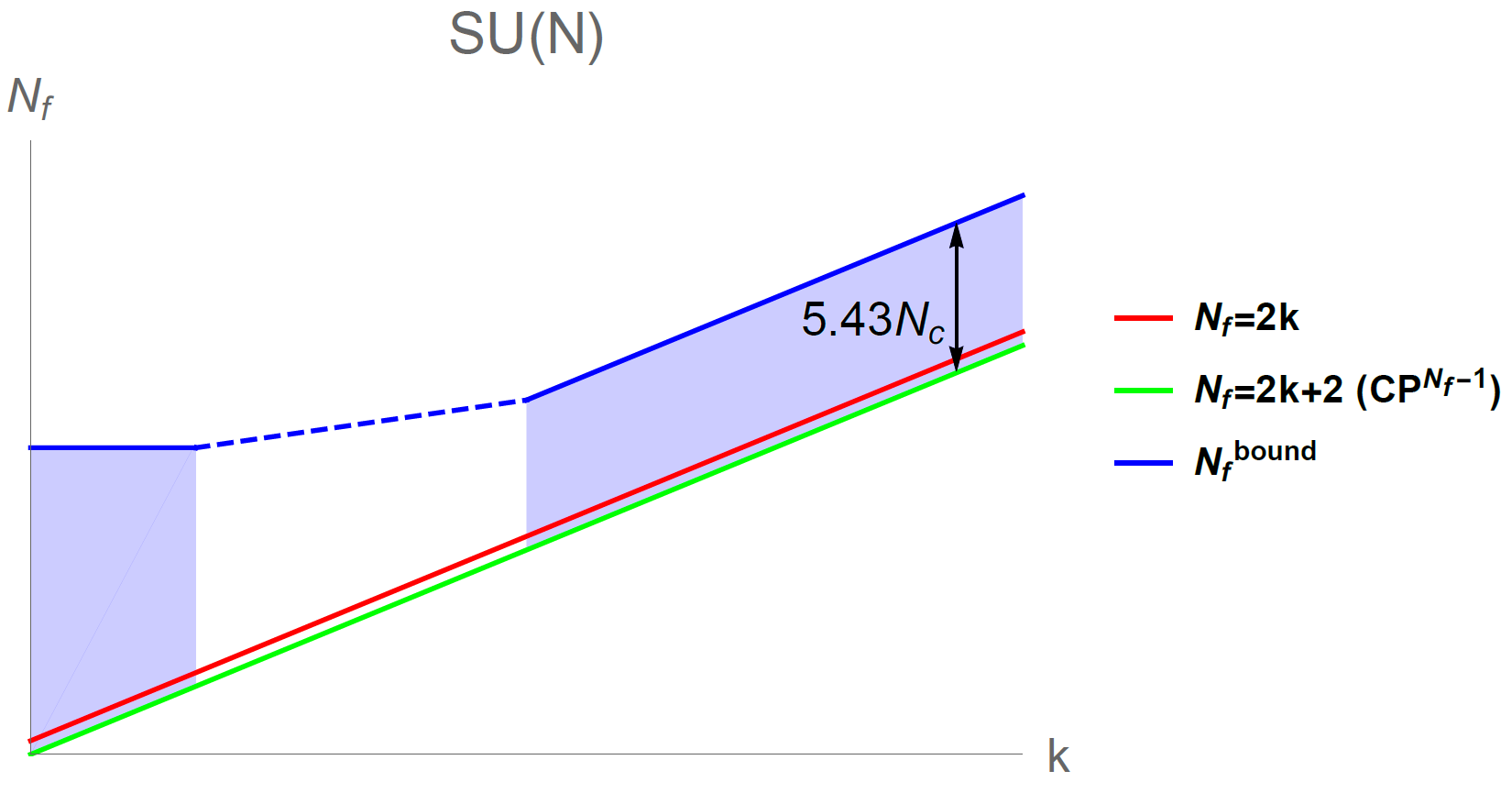}
	\captionsetup{width=.8\linewidth}
	\caption{General form of the results for $ N_f^{bound} $ for a general $ SU(N) $ gauge theory. The symmetry breaking phase can appear only inside the shaded area.}
	\label{fig:SU(N)}
\end{figure}

The purpose of the figure is to convey the general idea of the results, and it does not reflect any numerical calculations. The results are due to a saddle point approximation, but since the corresponding approximation for $ SU(2) $ led to excellent results, we expect the approximation here to be good as well.

The figure has three parts. The shaded area on the left corresponds to the results for $ k\ll N_f $, the shaded area on the right corresponds to large $ N_f $ and $ k\lesssim \frac{N_f}{2} $, and the unshaded area for intermediate $ k $ corresponds to $ k $'s for which neither approximation is useful. 

The results here are similar to those discussed in Section \ref{conclusions} for an $ SU(2) $ gauge theory. In particular, we once again find that we cannot exclude a finite-sized symmetry-breaking window of size $ 5.43N $ when $k\rightarrow \infty$. As discussed in Section \ref{conclusions}, we expect the size of the window to approach zero for large $ k $, and the fact that it has finite size might be a result of the large amount of fields we must integrate out in flowing from $ SQCD_3 $ to $ QCD_3 $. The results obtained here agree with this argument, since for large $ N_f $ we must integrate out about $ N\cdot N_f $ fields in flowing from $ SQCD_3 $ to $ QCD_3 $, and so we expect the bound to be weaker as $ N $ increases. We also once again find an agreement with the fact that the size of the symmetry breaking window must be maximized at $k=0$, and that the slope of $N_*(k)$ must be at most $2$ (as discussed in Section \ref{conclusions}).

\section{Summary and Discussion}

In this paper we used the method described in \cite{Grover2012} in order to bound the possible parameter space in which the symmetry-breaking phase conjectured in \cite{Komargodski2017} can occur. In particular, assuming that symmetry breaking can occur for $ 2k<N_f<N_\star $, we find a bound $ N_f^{bound} $ such that $ N_\star \leq N_f^{bound} $. For an $SU(2)$ gauge theory, we find exact bounds $ N_f^{bound} $, while for a general $SU(N)$ gauge theory the bounds  obtained here are only approximate, usually given in an expansion in $ \frac{1}{N_f} $ to order $ O(1) $. However, since the parameter $ N_f $ is discrete, these bounds can be considered precise for large enough $ N_f $ (i.e. when the corrections to $ N_f^{bound} $ are small enough). We also found that our results are comparable to lattice simulations results for the case $SU(2)_0$.

We find that we cannot exclude a symmetry-breaking window for any $ k $, and so our results support  the proposal in \cite{Komargodski2017}. The fact that we cannot exclude the symmetry-breaking phase even at large $ k $ is interesting, and there are two possible explanations for this fact. The more plausible explanation is that the window does indeed vanish for large enough $ k $, but our bound is not stringent enough to see this. Another explanation is that the symmetry-breaking phase persists to very large $ k $, which would be a very surprising result, since at large $ k $ the theory is weakly coupled. One way to find which of the two solutions is correct is to use a different RG flow. Indeed, we saw that starting with $ SQCD_3 $, one has to integrate out at least $ N\cdot N_f $ fields in order to flow to $ QCD_3 $. Since we worked in large $ N_f $, we thus expect the difference between $ F_{SQCD} $ and $ F_{QCD} $ to be quite large, leading to the resulting bound being weak. If one starts with a different theory whose F-coefficient is closer to $ F_{QCD} $, the resulting bound should be more stringent.

We conclude by noting that the recent developments in 2+1$d$ QFTs are likely to result in proposals for more dualities. Many dualities cannot be rigorously proven, and instead rely on various consistency checks, like the one described above. Unfortunately, while the bounds obtained above are rigorous, they are not ideal. A method with a "shorter" RG flow should lead to much better bounds in the theory discussed above, and also to better bounds in other examples. Hopefully this method will find many more uses in the future in which more stringent bounds will be obtained.

\section*{Acknowledgements}

The author would like to thank Z. Komargodski for the idea behind this paper and for many helpful discussions. The author would also like to thank R. Yacoby for helpful discussions and the Simons Center for Geometry and Physics for its generous hospitality. The author is supported by an Israel Science Foundation center for excellence grant and by the I-CORE program
of the Planning and Budgeting Committee and the Israel Science Foundation (grant
number 1937/12).

\newpage
\begin{appendices}
	
\addtocontents{toc}{\setcounter{tocdepth}{1}}
	
\section{Some F-Coefficients}\label{Some F-coefficients}

This appendix is a collection of F-coefficients used in the paper.

\subsection{Free Matter Fields}
The F-coefficient of an $ \mc{N}=2 $ free chiral multiplet is $$F=\frac{1}{2}\ln2$$
The F-coefficients of a free boson and a free Dirac fermion are (see \cite{Klebanov2011}):
\begin{align*}
F_\grf^{(d=3)}=& \frac{1}{2^4}\left(2\ln2-\frac{3\zeta(3)}{\pi^2}\right)\approx0.0638\\
F_\gry^{(d=3)}=& \frac{1}{2^3}\left(2\ln2+\frac{3\zeta(3)}{\pi^2}\right)\approx0.22
\end{align*}
Note how (as expected) the F-coefficient of a single $ \mc{N}=2 $ chiral multiplet is the same as  $ 2 F_\grf^{(d=3)} + F_\gry^{(d=3)}$.
\subsection{Gauge theories}
The F-coefficient of $ U(1)_k $ is $$F=\frac{1}{2}\ln k$$
The F-coefficient of pure $ \mc{N}=2 $ $ SU(2)_{\tilde{k}} $ is $$ F=-\log\left(\sqrt{\frac{2}{\tilde{k}}}\left|\sin\left(\frac{\pi}{\tilde{k}}\right)\right|\right) $$ 
This can be proven by explicitly calculating the integral \eqref{SU(2) partition function}. Note that when $ \tilde{k} $ is large enough\footnote{Specifically, we need $\tilde{k}\geq N$, so that we are not in the SUSY breaking phase.}, one can integrate out the adjoint fermion in the $ \mc{N}=2 $ vector multiplet (since it has a mass proportional to $ \tilde{k} $) and obtain non-SUSY $ SU(2)_{\tilde{k}-2} $, and so the results for the two theories should agree. Indeed, one can check that the results agree for all $ \tilde{k}\geq2 $  by comparing to \cite{Witten1988}.

\section{Saddle Point Approximation}\label{saddle_point_appendix}

\subsection{General Idea}\label{general idea appendix}
Assume we have some integral of the form
$$ I(k,N_f)= 2 \int dx e^{2 i \pi K x^2} e^{N_f G(x)}f(x) $$
where $ G $ is some function and $ K= a N_f + b $. We can thus write
$$ I(k,N_f)= 2 \int dx e^{N_f g(x)} e^{2 i \pi b x^2}f(x)  $$
where $ g(x)= 2 i \pi a x^2 + G(x)$. We can now perform a saddle point approximation when $ N_f $ is large. Define $ y=\sqrt{N_f}(x-x_0) $, where $ x_0 $ maximizes $ g(x) $. Taking $ N_f $ to be large, we can write
$$ e^{N_fg(x)}=e^{N_fg(x_0)}e^{y^2 g''(x_0)/2}\left(1+\frac{y^3g'''(x_0)}{6\sqrt{N_f}} + \frac{3y^4g''''(x_0)+y^6 (g'''(x_0))^2}{72N_f}+...\right)$$
In this paper, we will almost always have $ x_0=0 $\footnote{In parts of the paper, we will have $ x_0=O(1/N_f) $. When this occurs, we explain why this correction can be ignored to the order in $ N_f $ we will be working in, and so the above will still be valid.}. We thus plug in $ x_0=0 $ and obtain
$$ I(k,N_f)= \frac{2e^{N_fg(0)}}{\sqrt{N_f}} \int dy e^{y^2 g''(0)/2}\left(1+\frac{y^3g'''(0)}{6\sqrt{N_f}} + \frac{3y^4g''''(0)+y^6 (g'''(0))^2}{72N_f}+...\right) e^{2 i \pi b \frac{y^2}{N_f}}f\left(\frac{y}{\sqrt{N_f}}\right) $$
Which allows an expansion in $ N_f $:
$$ I(k,N_f)= \frac{2e^{N_fg(0)}}{\sqrt{N_f}} \int dy e^{y^2 g''(0)/2}\left(1+\frac{y^3g'''(0)}{6\sqrt{N_f}} + \frac{3y^4g''''(0)+y^6 (g'''(0))^2}{72N_f}+...\right) \left(1+2 i \pi b \frac{y^2}{N_f} + ...\right)\left(f(0)+f'(0)\frac{y}{\sqrt{N_f}}+...\right)$$
Collecting powers of $ N_f $ will then give the desired result for $ I(k,N_f) $ as an expansion in $ N_f $.

\subsection{SU(2) F-coefficients Without F-maximization Using the Saddle-Point Approximation}\label{SU(2)_without F_max_appendix}

The integral we have to calculate is equation \eqref{SU(2) partition function}, with $ \grD=\frac{1}{2} $:
$$
Z_{S^3}(k,N_f) = 2 \int dx e^{-2 i \pi K x^2} e^{ N_f (l(1/2 +ix)+l(1/2 -ix))} \sinh^2(2 \pi x)
$$
From which we can obtain the F-coefficient by calculating
$$F = -\ln |Z_{S^3}|$$
which we can expand in powers of $ \frac{1}{N_f} $.

Using the fact that $ e^{l(1/2 +ix)+l(1/2 -ix)}=\frac{1}{2\cosh(\pi\grl)} $ \cite{Hosomichi2010}, we can simplify the integral and write it as
$$
Z(k,N_f) = \frac{1}{2^{N_f-1}} \int dy e^{-2i \pi K x^2}\frac{\sinh^2(2 \pi x)}{\cosh^{N_f}(\pi x)}
$$
Performing the saddle point approximation as explained in Appendix \ref{general idea appendix}, we obtain
$$ Z=2^{1-N_f}\int dx\left( 4 \pi ^2  y^2 e^{\frac{1}{2} \left(-\pi ^2-4 i \pi  a\right) y^2}\left(\frac{1}{N_f}\right)^{3/2}+O\left(\frac{1}{N_f}\right)^{5/2}\right) $$
and so $$ F = N_f \ln2 + \frac{3}{2}\ln N_f + \frac{1}{2}\ln\left(\frac{(16a^2 + \pi^2)^{3/2}}{128\pi^2}\right) + O\left(\frac{1}{N_f}\right) $$
The first two terms in this expression can be understood intuitively, and will be explained in a more general context in Section \ref{Nstar for SU(N)}. This result can also be compared with the result for the $ U(N) $ case in \cite{Klebanov2011_2}. Under the correct replacements which make it compatible with an $ SU(N) $ gauge theory, the two results agree.

\end{appendices}
\pagebreak

\bibliographystyle{utphys}
\providecommand{\href}[2]{#2}\begingroup\raggedright\endgroup

\end{document}